# On the Impact of Partial Sums on Interconnect Bandwidth and Memory Accesses in a DNN Accelerator

Mahesh Chandra
NXP Semiconductors, India

*Abstract*— Dedicated accelerators are being designed to address the huge resource requirement of the deep neural network (DNN) applications. The power, performance and area (PPA) constraints limit the number of MACs available in these accelerators. The convolution layers which require huge number of MACs are often partitioned into multiple iterative sub-tasks. This puts huge pressure on the available system resources such as interconnect and memory bandwidth. The optimal partitioning of the feature maps for these sub-tasks can reduce the bandwidth requirement substantially. Some accelerators avoid off-chip or interconnect transfers by implementing local memories; however, the memory accesses are still performed and a reduced bandwidth can help in saving power in such architectures. In this paper, we propose a first order analytical method to partition the feature maps for optimal bandwidth and evaluate the impact of such partitioning on the bandwidth. This bandwidth can be saved by designing an active memory controller which can perform basic arithmetic operations. It is shown that the optimal partitioning and active memory controller can achieve up to 40% bandwidth reduction.

## I. INTRODUCTION

Deep neural networks are used for the state of the art results in multiple applications. The convolution layers used in these networks require huge computing power and data movement. To achieve real time performance, dedicated accelerator architectures are being explored and designed [1,2,3]. However, due to cost, technology and power constraints, number of MACs are still limited in these accelerators. Often, these are not enough to compute all the output feature maps or process all the input feature maps for many layers. So, there is a requirement of partitioning the convolution task into multiple sub tasks and instead of processing all the input or output maps, only some of them are processed in one iteration.

Since, the CNNs come with different flavors, effective PE utilization is one of the key issues with the DNN accelerators. PE utilization is affected by data flow and partitioning of convolution tasks. Loop optimization techniques such as loop unrolling, loop tiling and loop interchange are used to find the right configuration for the efficient data flow and minimum data movement[4]. Different data reuse strategy are adopted to optimally use the system resources and architect the accelerators [5].

The strategies used for reusing the weights, input activations or output activations improve the PE utilization and reduce the amount of data transfer. However, each of these strategy has a different impact on the data bandwidth requirement due to the input maps, output maps and weights. These strategy also often require the partial sums to be generated as all channels can't be accommodates in PEs. These partial sums double the required bandwidth because the partial sums must be read before being updated. Please note that we use the word bandwidth, in case of local memories, it'll translate to memory accesses. In case of local memories, it's a big factor for power consumption, so reducing it is as important as the off-chip or interconnect bandwidth. Though, there is general agreement on the increased bandwidth requirement due to partial sums, a quantitative analysis is missing. This analysis for multiple CNNs is reported in section IV of this paper.

The CNNs are feed forward neural networks. Activation of a layer can be computed only when the activations of previous layer or layers are available. So, once a loop tiling approach is decided, we can assume that all the activation for that particular layer will be computed before the next one. So far, the criteria is mainly the PE utilization for partitioning the input and output channels. As the partial sums require lot of bandwidth, attention must also be paid to BW while partitioning the channels. The bandwidth requirement varies with the partitioning of the input and output feature maps. An optimal partitioning is thus key to the bandwidth increase in an accelerator. One such method is discussed in section II of this paper.

The DNNs require both compute power and data transport. Researchers are working on processing near memory and processing in memory [6]. Processor in memory, intelligent RAM and other such architectures integrate the processors within the memory itself. These architectures will have huge impact on the future DNN accelerators. However, for the current set of accelerators, some simple solution such as active memory controller [7] have huge impact. This paper also discusses the impact of the active memory controller on the bandwidth. The off-chip data transfers results in high power consumption. So, the accelerators are designed with more on-chip memory. The proposed active memory controller can also be helpful in optimizing bandwidth in such accelerators and hence, reduce the power consumption.

## II. METHOD OF OPTIMALLY PARTITIONING THE FEATURE MAPS

The code for a convolution layer in a CNN, processing M input feature maps of size $W_i \times H_i$ and generating N output maps of size $W_o \times H_o$ using a kernel KxK, can be written as:

```
for (co=0;co<N;co++)
    for (ci=0;ci<M;ci++)
        for (x=0;x<Wi;x++)
            for (y=0;y< Hi;y++)
                for (k=-K;k<=K;k++)
                    for (l=-K;l<=K,l++)
                        f_out[co][x][y] += f_in[ci][x+k][y+l] * wt[co][ci][k][l];
```

For maximum data reuse, we'd ideally like to process all the input map only once and compute the output map at one iteration as more often they are the highest contributors to the data bandwidth. However, that requires a large number of MACs in the accelerator. In practice, we have limited number of MAC units in the accelerator and so, the input and output maps are partitioned and only partial sums are computed. It means that computing an output feature map requires multiple iteration. Similarly, an input map may be read more than once for computing different output maps. If we process n input maps and m output maps on one iteration, the code sequence above is written as:

*for (co_base=0;co_base< N;co_base=co+n)*

  *for (co=0;co<n;co++)*

    *for (ci_base=0;ci_base<M;ci_base=ci_base+m)*

      *for (ci=0;ci<m;ci++)*

        *for (x=0;x<Wi;x++)*

          *for (y=0;y< Hi;y++) {*

            *for (k=-K;k<=K;k++)*

              *for (l=-K;l<=K;l++)*

                *p_sum[co] += f_in[ci_base+ci][x+k][y+l] * wt[co_base +co][ ci_base +ci][k][l];*

              *f_out[co_base+co][x][y] = f_out[co_base+co][x][y] + p_sum[co];}*

The choice of *m* and *n* trades-offs the data reuse at input and output. In this section, we discuss the first order optimal selection of *m* and *n*.

Let's first make some assumptions for the first order model. Let the number of MACs in the accelerator be *P*. So, these *P* multipliers, that the accelerator can handle, need to be divided among the input and output maps for computing partial sums. Any given choice of *m* and *n* must satisfy:

$$K^2 \times m \times n < P \qquad (1)$$

The input map will be read *N/n* times and output map (partial sums) is written *M/m* times. The output map will also be read before being updated for all the iteration except the first one. So, the Input and output BWs ($B_i$ and $B_o$ respectively here) are given by:

$$B_i = W_i \times H_i \times M \times \frac{N}{n} \qquad (2)$$

$$B_o = W_o \times H_o \times N \times (2 \times \frac{M}{m} - 1) \qquad (3)$$

Total bandwidth (in terms of million activation per inference) is given by:

$$B = B_i + B_o = W_i \times H_i \times M \times \frac{N}{n} + W_o \times H_o \times N \times (2 \times \frac{M}{m} - 1) \qquad (4)$$

For the purpose of this model, let's assume that we can find m and n such that

$$K^2 \times m \times n = P \qquad (5)$$

The eq (4) can be now written as:

$$B(m) = W_i \times H_i \times M \times \frac{N}{P} \times K^2 \times m + W_o \times H_o \times N \times (2 \times \frac{M}{m} - 1) \qquad (6)$$

To minimize, B(m),

$$\frac{dB}{dm} = 0$$

Which results in

$$m = \sqrt{\frac{2 \times W_o \times H_o \times P}{W_i \times H_i \times K^2}} \qquad (7)$$

Value of *m* obtained by eq. (7) is a real number and cannot be used directly; so, it must be adapted. For the purpose of this analysis, the value of m is slightly modified so that it is integer and it is a factor of *M* (i.e. total number of input channels in the layer). Once *m* is known, *n* can be computed from eq (5).

III. REDUCING BW USING ACTIVE SRAM CONTROLLER

As discussed above, to update the result of convolution, the accelerator must read the previous partial sum before adding new one and writing it to memory. So, each operation, except the first one, requires a read and a write operation. Active SRAM controller have been proposed in past to offload some of the CPU tasks in case of data-heavy processing [7]. Similar to these active controller, if we design a memory controller which is capable of doing simple operations like addition, initialization and compare etc., we can avoid reading of the data from the memory and save lot of memory accesses and data bandwidth. There are two main challenges, one, which operations to offload because the memory controller can become as complex as compute engine if care is not taken and two, how to communicate to memory controller about the type of operation. We address these two issues in this section.

The working principle is that instead of transporting data (i.e. previous partial sums, in this case) all the way to the compute engine (or MAC units), perform these operations (i.e. the addition) locally in memory controller. MACs compute the partial sum and communicate to the memory controller to update the memory location instead of writing with the value provided by the compute engine (MAC block). The memory controller will perform a read-update-write operation. We require following operation quite often and handling would reduce the bandwidth requirement significantly:

- **Addition**. The partial sums computed by compute engine needs to be added to previously stored partial sums for all but the first iteration.

- **Activation**. The last partial sum update is often followed by activation unit. However, the scaling may be needed before the activation. Some activation functions like ReLu are quite simple and can be integrated within memory controller. This can also help in offloading the compute engine and save bandwidth. If more than one activation function is supported by memory controller, the selection can be done statically by programming appropriate configuration registers.

- **Normal**. The normal memory read/write mode.

The second problem is to communicate with the SRAM controller. Note that the DNNs require huge data bandwidth and would require the latest advanced bus interface e.g. AXI4. These advanced interfaces support the sideband signals which are part of the infrastructure and travel through the interconnect. For example, AXI4 has the '*awuser*' signals [8] which can be used for signaling the command to the SRAM controller. The representative block diagram is shown in the figure 1.

TABLE I. BANDWIDTH (MILLION ACTIVATIONS/IMAGE) REQUIREMENT DIFFERENT PARTITIONING STRATEGIES AND NUMBER OF MACs (P) IN ACCELERATOR

| CNN | P=512 | | | | P=2048 | | | | P=16384 | | | |
|---|---|---|---|---|---|---|---|---|---|---|---|---|
| | Max Input | Max Output | Equal MACs | This Work | Max Input | Max Output | Equal MACs | This Work | Max Input | Max Output | Equal MACs | This Work |
| AlexNet | 61.9 | 94.2 | 26.2 | *25.1* | 52.2 | 64.6 | 13.0 | *12.6* | 9.2 | 10.9 | 7.3 | *4.3* |
| VGG-16 | 1170.3 | 1938.6 | 494.2 | *442.5* | 909.5 | 1309.3 | 269.3 | *237.2* | 207.1 | 241.1 | 151.0 | *83.5* |
| SqueezeNet | 199.6 | 244.8 | 65.9 | *52.0* | 53.6 | 105.2 | 47.4 | *26.2* | 12.6 | 17.3 | 34.8 | *11.1* |
| GoogleNet | 431.7 | 313.6 | 102.5 | *93.5* | 174.6 | 151.6 | 61.2 | *47.7* | 23.8 | 24.1 | 41.6 | *17.5* |
| ResNet-18 | 281.2 | 315.8 | 96.1 | *88.9* | 205.0 | 191.6 | 50.9 | *46.8* | 35.1 | 31.7 | 26.9 | *16.0* |
| ResNet-50 | 5245.2 | 5770.4 | 1059.2 | *952.6* | 2909.0 | 2830.4 | 608.6 | *479.5* | 929.8 | 682.5 | 330.1 | *168.5* |
| MobileNet | 215.0 | 209.2 | 78.5 | *68.3* | 136.8 | 116.2 | 48.8 | *35.0* | 21.9 | 21.0 | 34.9 | *16.1* |
| MNASNet | 884.4 | 1294.1 | 405.3 | *373.4* | 722.0 | 1030.3 | 213.4 | *183.0* | 500.2 | 516.3 | 101.8 | *66.0* |

TABLE II. BANDWIDTH (MILLION ACTIVATIONS/IMAGE) REQUIREMENT FOR PASSIVE AND ACTIVE MEMORY CONTROLLER DEPENDING ON NUMBER OF MACs (P) IN THE ACCELERATOR

| CNN | Passive Memory Controller | | | | | | Active Memory Controller | | | | | |
|---|---|---|---|---|---|---|---|---|---|---|---|---|
| | 512 MACs | 1024 MACs | 2048 MACs | 4096 MACs | 8192 MACs | 16384 MACs | 512 MACs | 1024 MACs | 2048 MACs | 4096 MACs | 8192 MACs | 16384 MACs |
| AlexNet | 25.07 | 17.54 | 12.56 | 8.89 | 6.52 | 4.32 | 17.89 | 12.62 | 8.77 | 6.38 | 4.55 | 3.51 |
| VGG-16 | 442.49 | 321.79 | 237.25 | 169.43 | 112.14 | 83.54 | 315.33 | 225.44 | 161.67 | 123.36 | 89.97 | 63.67 |
| SqueezeNet | 51.98 | 37.47 | 26.22 | 20.04 | 14.12 | 11.10 | 40.06 | 27.35 | 20.76 | 14.87 | 12.61 | 9.78 |
| GoogleNet | 93.46 | 67.17 | 47.65 | 35.20 | 23.23 | 17.51 | 69.90 | 48.37 | 35.77 | 25.95 | 20.63 | 14.62 |
| ResNet-18 | 88.87 | 63.56 | 46.79 | 32.86 | 22.01 | 16.02 | 63.52 | 45.53 | 32.34 | 24.74 | 17.81 | 12.90 |
| ResNet-50 | 952.60 | 691.13 | 479.50 | 349.75 | 232.82 | 168.46 | 691.98 | 480.49 | 346.77 | 242.90 | 183.09 | 121.93 |
| MobileNet | 68.53 | 46.74 | 35.14 | 25.22 | 21.00 | 16.02 | 50.90 | 39.03 | 27.69 | 22.66 | 17.82 | 15.58 |
| MNASNet | 373.41 | 264.36 | 183.01 | 128.27 | 92.35 | 65.96 | 258.91 | 188.75 | 131.06 | 94.92 | 67.80 | 50.40 |

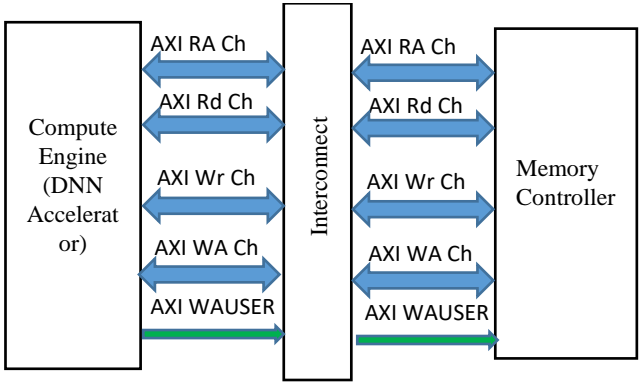

Fig. 1. Communication between compute engine and memory controller through using advanced bus protocols

For the purpose of analysis in this paper, only offloading of the addition of partial sums to the memory controller is assumed. It results in significant reduction of bandwidth as discussed in next section.

## IV. RESULTS AND DISCUSSION

We first evaluated the minimum bandwidth required for some of the popular CNNs. Table III lists the cumulative bandwidth required by the convolution layers of some of the popular CNNs, assuming that the output of convolution layer is written to the memory (i.e. no fused operations across layers) and the data is read and written only once (i.e. we have unlimited compute resources and don't need to write partial sums). Note that the first assumption also applies to the rest of the discussion in this section. This bandwidth is computed for the inference on a 224x224 color image.

TABLE III. MINIMUM BW REQUIREMENT FOR CNNs

| CNN | BW (M Activations/inference) |
|---|---|
| AlexNet [9] | 0.823 |
| VGG-16 [10] | 20.095 |
| SqueezeNet [11] | 7.304 |
| GoogleNet [12] | 7.889 |
| Resnet-18 [13] | 4.666 |
| Resnet-50 [13] | 28.349 |
| MobileNet [14] | 10.273 |
| MNASNet [15] | 11.001 |

In a constrained system, the bandwidth impact is decided by the partitioning of input and output maps. Table I compare the bandwidth requirements for the input and output maps partitioning using four different method for three different MAC numbers. These methods are:

1. Allocate the MACs to maximum number of input maps. This will reduce the number of output iterations. This is reported in 1st columns in table I.

2. Allocate the MACs to maximum number of output maps. This will reduce the number of input iterations. This is reported in 2nd columns in table I.
3. Allocate equal number of MACs to input and output channels. This is reported in 3rd columns in table I.
4. Allocate as per the first order optimum number as presented in this paper in section II. This is reported in 4th columns in table I.

Note that the method presented here results in the minimum bandwidth compared to all others. Also note that as number of MACs increases, the required bandwidth decreases and with a very large number of MACs, it approaches the minimum bandwidth as given in table III. So, it is obvious that the partitioning strategy is very critical in resource constrained systems such as IoT and low power cores.

If the SRAM controller is updated for performing additions locally as discussed in section III, the required bandwidth is significantly reduced. The absolute numbers are given in table II and percentage reduction is shown in figure 2.

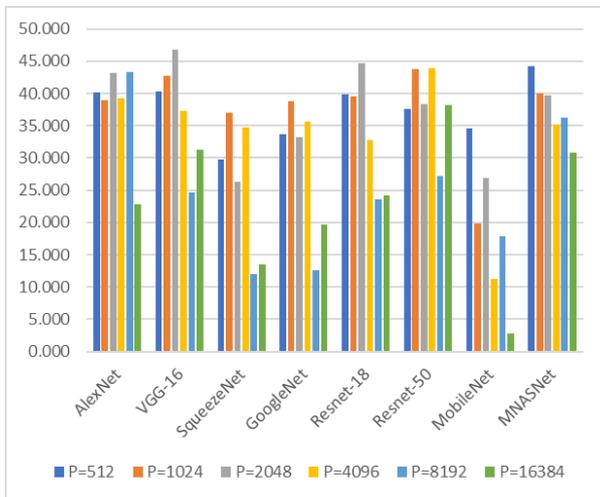

Fig. 2. Percentage bandwidth saving with active SRAM controller

As the number of MACs in accelerators is increased, the gain from active SRAM controller is reduced. However, even with relative huge number of MACs (16K) in the device, the performance gain varies between 2-38% which is significant. As expected the gain is significantly higher at 19-42% for more constrained compute systems (i.e. less number of MACs).

From the data above, it is obvious that huge bandwidth can be saved using the active memory controller. The cost of supporting this is very small logic in memory controller and handling the sideband signals on the interconnect. This cost is almost negligible. An obvious advantage of reducing this bandwidth is reduced power consumption.

V. CONCLUSION

In this paper, the impact of the partial sum computation on the memory bandwidth is evaluated. A method for partitioning the feature maps is proposed that helps in reducing the bandwidth significantly. Active memory controller can further reduce the bandwidth requirement, This approach can be useful in designing the machine learning accelerators. The methods presented can be applied for low power AI systems as they reduce the number of memory accesses and data transport.

REFERENCES

[1] Y. Chen, Y. Xie, L. Song, F. Chen, and T. Tang, "A Survey of Accelerator Architectures for Deep Neural Networks", Engineering, 2020
[2] J. Wang, J. Lin and Z. Wang, "Efficient Hardware Architectures for Deep Convolutional Neural Network," IEEE Transactions on Circuits and Systems I: Regular Papers, 65(6), 2018
[3] V. Sze, Y. Chen, J. Emer, A. Suleiman and Z. Zhang, "Hardware for machine learning: Challenges and opportunities", 2017 IEEE Custom Integrated Circuits Conference (CICC), Austin, TX, 2017
[4] Y. Ma, Y. Cao, S. Vrudhula, and J. Seo, "Optimizing loop operation and dataflow in FPGA acceleration of deep convolutional neural networks." In Proceedings of the 2017 ACM/SIGDA International Symposium on Field-Programmable Gate Arrays, 2017.
[5] Y. Chen, J. Emer and V. Sze, "Using Dataflow to Optimize Energy Efficiency of Deep Neural Network Accelerators", IEEE Micro, 37(3), 2017
[6] S. Ghose, K. Hsieh, A. Boroumand, and R. Ausavarungnirun,"Enabling the Adoption of Processing-in-Memory: Challenges, Mechanisms, Future Research Directions", arXiv:1802.00320v1, 2018
[7] J. Yoo, S. Yoo, and K. Choi, "Active Memory Processor for Network-on-Chip-Based Architecture", IEEE Trans. on computers, 61(5), 2012
[8] ARM, "Axi4-stream protocol specification"
[9] A. Krizhevsky, I. Sutskever, and G. E. Hinton, "ImageNet classification with deep convolutional neural networks" NIPS, 2012
[10] K. Simonyan, and A. Zisserman, " deep convolutional networks for large-scale image recognition", arXiv:1409.1556, 2014
[11] F. N. Iandola, S. Han, M. W. Moskewicz, K. Ashraf, W. J. Dally, and K. Keutzer, "Squeezenet: Alexnet-level accuracy with 50x fewer parameters an <0.5MB model size", arXiv:1602.07360v4, 2016
[12] C. Szegedy,W. Liu, Y. Jia, P. Sermanet, S. Reed, D. Anguelov, D. Erhan, V. Vanhoucke, and A. Rabinovich, "Going deeper with convolutions." Proceedings of the IEEE CVPR, 2015
[13] K. He, X. Zhang, S. Ren, And J. Sun, "Deep residual learning for image recognition", Proceedings of the IEEE CVPR, 2016
[14] M. Sandler, A. Howard, M. Zhu, A. Zhmoginov, and L. C. Chen, "Mobilenetv2: Inverted residuals and linear bottlenecks", In Proceedings of the IEEE CVPR, 2018
[15] M. Tan, B. Chen, R. Pang, V. Vasudevan, M. Sandler, A. Howard, and Q. V. Le, "Mnasnet: Platform-aware neural architecture search for mobile", In Proceedings of the IEEE CVPR, 2019